\documentclass[prb,showpacs,preprint,amsmath,amssymb,groupadress,superscriptaddress]{revtex4}
\usepackage{graphicx}
\begin{document}

\title{Symmetry constraints on the electrical polarization in novel multiferroic materials}

\author{P.G. Radaelli}
\affiliation{ISIS facility, Rutherford Appleton Laboratory-CCLRC,
Chilton, Didcot, Oxfordshire, OX11 0QX, United Kingdom. }
\affiliation{Dept. of Physics and Astronomy, University College
London, Gower Street, London WC1E 6BT, United Kingdom}
\author{L.C. Chapon}
\affiliation{ISIS facility, Rutherford Appleton Laboratory-CCLRC,
Chilton, Didcot, Oxfordshire, OX11 0QX, United Kingdom. }
\date{\today}

\begin{abstract}
The symmetry conditions for the development of a macroscopic electrical polarization as a secondary order parameter to a magnetic ordering transition, and the constraints on the direction of the polarization vector, are determined by a non-conventional application of the theory of irreducible co-representations.  In our approach, which is suitable for both magnetic and structural modulations, anti-unitary operators are employed to describe symmetry operations that exchange the propagation vector $\textbf{k}$ with $\textbf{-k}$, rather than operations combined with time-reversal as in classical \textit{corep} analysis. Unlike the conventional irreducible representations, co-representations can capture the full symmetry properties of the system even if the propagation vector is in the interior of the Brillouin zone.  It is shown that ferroelectricity can develop even for a completely collinear structure, and that helical and cycloidal magnetic structures are not always polar. In some cases, symmetry allows the development of polarization parallel to the magnetic propagation vector. Our analysis also highlights the unique importance of magnetic commensurability, enabling one to derive the different symmetry properties of equivalent commensurate and incommensurate phases even for a completely generic propagation vector.
\end{abstract}


\maketitle

\section{Introduction}

There has been a recent surge of interest for improper ferroelectric transition-metal compounds where the onset of electrical polarization is inducved by a transition to a complex magnetic state \cite{Kimura_nature_2003}. Much of the discussion in the literature has focused on establishing the microscopic mechanism that couples the magnetic moments with lattice displacements: symmetric superexchange (sometime referred to as "superexchange striction")  and antisymmetric exchange (the so-called Dzyaloshinskii-Moriya contribution) have both been discussed, often in the context of the same materials \cite{Sergienko1, Sergienko2}, and shown in many cases to describe the details of the coupled magneto-electric transitions, including the change in sign or direction of the polarization in an applied magnetic field.  The study of the crystal and magnetic symmetry of these systems represents an integral part of this work. Establishing the symmetry constraints upon the polarization vector for a given magnetic structure, regardless of the microscopic mechanism of magneto-elastic coupling, is extremely valuable. For example, if it is established that for a given magnetic structure the polarization vector $\textbf{P}$ lies by symmetry along a particular crystallographic direction, the experimental determination of the direction of $\textbf{P}$ can be used to corroborate or falsify the magnetic structure model.  Conversely, the exact direction of $\textbf{P}$ must be predictable from the microscopic coupling mechanism when symmetry allows $\textbf{P}$ to lie in a plane or in a general direction. More generally, and in analogy with the well-know case of magnetostriction, if the broken symmetry allows the development of a macroscopic polar vector, one would \emph{always} expect to observe ferroelectricity, provided that the measurement has sufficiently high sensitivity.  The "new" class of multiferroic materials we are discussing is in fact characterized by very small values of $|\textbf{P}|$. However, the \emph{magnitude} of $\textbf{P}$ must be predictable from microscopic models based on the specific observed spin arrangement, out of the many that are usually consistent with each symmetry class.

A number of techniques exist to predict possible symmetries derived from a given parent structure through a given order parameter - a subject that was thoroughly developed in the 1980's \cite{Toledano}.  When the propagation vector of the structural or magnetic modulation lies in a high-symmetry point of the Brillouin zone, one can use standard irreducible representations \cite{Bertaut_1962, Bertaut_1968, Bertaut_1971, Bertaut_1981} (\textit{irreps}) to generate the so-called \emph{image} of the high-symmetry group - the finite set of matrices representing the group elements in the order parameter space for a given \textit{irrep}.  By analyzing the image one can determine the form of the Landau free energy as a function of both primary and secondary order parameters, enumerate the possible invariance groups of the low-symmetry phase and establish compliance with the so-called Landau criterion for continuous phase transitions \cite{Toledano}.  When dealing with magnetic transitions, some of the early work employed irreducible \emph{co-representations} (\textit{coreps}) instead of \textit{irreps}.  In this approach, first discussed by Wigner \cite{Wigner_book}, the time reversal operator in the black-and-white symmetry groups is \emph{antiunitary}, as applied to the Schrodinger wavefunction describing the magnetic ground state. However, it was later noted (see for example \cite{Izyumov}) that the use of antiunitary operators is not actually necessary when dealing with classical spin systems, for which time reversal is represented by a simple sign change, and the use of \textit{coreps} has been all but abandoned as a consequence.

When the propagation vector of a magnetic or modulated crystal structure is not a high-symmetry point of the Brillouin zone, the situation is considerably more complex.  First of all, the "mathematical" or "physical" representations are in this case always obtained by combining at least two little-group-\textit{irreps} with opposite propagation vectors ($\textbf{+k}$ and $\textbf{-k}$).  Secondly, the image of this representation can have an infinite number of elements if the propagation vector is incommensurate with the crystal lattice.  Thirdly and perhaps more importantly, the symmetry properties of the modulated structure upon application of operators that exchange $\textbf{+k}$ with $\textbf{-k}$, such as the inversion, do not emerge clearly, since these operators do not have an image matrix in either of the \textit{irreps}.  For these reasons, image analysis is not ordinarily applied to these problems.  Instead, one usually resort to constructing Landau free energies containing mixed terms in the $\textbf{+k}$ and $\textbf{-k}$ order parameters, and examining the symmetry of the solutions.  Harris and collaborators have extensively employed this approach to study in some detail the specific case of incommensurate magnetic multiferroics \cite{Kenzelmann_2005, Lawes, Harris_jap, Kenzelmann_Ni3V2O8, Harris_review_2006, Harris_review2_2006}.  In particular, they have specifically pointed out the importance of inversion symmetry, the globally invariant forms of the Landau free energy and the situations where this symmetry is spontaneously broken, giving rise to a net polarization.

In this paper, we follow a different approach to the problem of determining the point-group symmetry of a modulated magnetic structure and of defining the symmetry conditions imposed on the development of a macroscopic polar vector upon magnetic ordering with an arbitrary propagation vector $\textbf{k}$.  In our analysis, we employ the mathematical tool of irreducible co-representations, but its significance is radically different from that of the "standard" \textit{corep} analysis of magnetic structures.  This difference is apparent when one considers the role of antiunitary operators, that are conventionally employed to represent "black" elements of the groups inverting the direction of time.  Here, we use antiunitary elements to describe operators that exchange $\textbf{+k}$ with $\textbf{-k}$, as explained below. This approach enables one to employ a \emph{single} \textit{corep} and propagation vector instead of two \textit{irreps}.  Time reversal is not an essential ingredient of this method, which can be equally well employed to study incommensurate \emph{structural} modulations (for magnetic structures, we treat time reversal classically, es explained below).  Crucially, all the symmetry properties emerge naturally from this analysis, since the "little group" of the propagation vector is extended to include the inversion and all other operators exchanging $\textbf{+k}$ with $\textbf{-k}$.  This approach is not particularly new - it is implicit in the treatment of \textit{coreps} described in the classic book by Kovalev \cite{Kovalev_1993} and its significance has been recently re-emphasized by Schweizer \cite{Schweizer_CR05}.  However, we do not believe that these techniques have been hitherto employed to determine crystallographic point groups and macroscopic observables, as we do herein.  With respect to the approach followed by Harris and coworkers, the main difference is that we perform our analysis directly on the \emph{images}, as in the case of high-symmetry \textbf{k} vectors.  Knowledge of the Landau free energy form is therefore not required, making this method easier to implement in an automated and tabulated form, as appropriate for non-specialists. We nonetheless stress that the Landau analysis provides much more information than the point group of the low-symmetry phase, and is therefore the tool of choice for specialist theoreticians. Furthermore, our method is completely general - inversion symmetry is treated on an equal footing with other operators exchanging $\textbf{+k}$ with $\textbf{-k}$.  Our only restriction is that the magnetic structure be described by a \emph{single} $\textbf{k}$, although the extension of our analysis to multi-$\textbf{k}$ structures is quite straightforward.

We show that ferroelectricity can develop even when the magnetic structure
is described by a single order parameter, and that $\textbf{P}\parallel\textbf{k}$ is allowed by symmetry in some cases. Furthermore, our analysis evidences the crucial difference between \emph{incommensurate} and \emph{commensurate} magnetic structures even for propagation vectors inside the Brillouin zone, and, for the latter, shows that the global phase has an influence on symmetry. This is particularly counterintuitive, since the global phase affects neither the magnetic energy nor the intensity of the magnetic Bragg peaks in (unpolarized) neutron diffraction. In fact, the observation of a non-zero electrical polarization can be used to discriminate between otherwise indistinguishable magnetic structures.

The paper is organized as follows:  in Section \ref{section:theory} we describe the use of \textit{corep} to determine the point-group symmetry of a magnetically modulated system (the extension to lattice modulations is straightforward, and will be described elsewhere).  In Section \ref{section: examples} we describe in detail the application of this method to a number of topical multiferroic systems with commensurate or incommensurate magnetic structures.  Section \ref{section: discussion} contains summary and discussion of the results.  For completeness, in Appendix \ref{Appendix_coreps} we provide a brief overview of the theory of irreducible co-representations.

\section{Theory}
\label{section:theory}

As for all macroscopic observables that are even by time reversal, the existence of an electrical polarization and the restrictions on its direction are defined by the structural point group $S$ of the magnetically ordered structure $m$. If we write a generic element of the paramagnetic space group $G$ in the form $g=\{r|\textbf{w}\}$ (Seitz notation), where $r$ is a proper or improper rotation belonging to the paramagnetic point group and $\textbf{w}$ is a translation, then $r \in S$ if and only if there is an element $g \in G$ for which $g m
= \pm m$. If we only consider the \emph{representative} elements of the group, $g_0 = \{r|\textbf{v}\}$, where $\textbf{v}$ is a non-Bravais translation and rotations appear only once in the representative set, then

\begin{equation}
\label{pg_condition} r \in S \longleftrightarrow \{r|\textbf{v}\} m
= \pm \{E|\textbf{t}\}m
\end{equation}

where $\textbf{t}$ is a Bravais translation and $E$ is the identity. In other words, the corresponding representative element must be equivalent to $\pm$ a lattice translation.  Here, we deal with the time reversal symmetry by considering both positive and negative eigenvalues, rather than combining time reversal and complex conjugation as it is often done \cite{Wigner_book, Bradley_Cracknell}. It can be shown that the two approaches are completely equivalent \cite{Izyumov}.  It is important at this point to recognize  that the magnetic structure $m$ is \emph{real}, i.e., it is not a generic element of the linear space $\textbf{V}$ defined over the complex field as a collection of axial vectors associated with atomic positions in the crystal. If $m$ is described by a single propagation vector $\textbf{k}$, we can always write $m=e^{+i\textbf{k}\cdot\textbf{t}}\psi+e^{-i\textbf{k}\cdot\textbf{t}}\psi^*$, where $e^{+i\textbf{k}\cdot\textbf{t}}\psi$ is a generic element of the complex-valued subspace of $\textbf{V}$ associated with the "arm" $\textbf{k}$.  In the most general case, $e^{+i\textbf{k}\cdot\textbf{t}}\psi$ and $e^{-i\textbf{k}\cdot\textbf{t}}\psi^*$ transform with distinct (albeit complex-conjugate) representations of $G$, and in fact may not even belong to the same star.  Conventional analysis using the \textit{irreps} of the little group $G_k$ (i.e., of the group of operators leaving the propagation vector invariant) deals with the two Fourier components separately, combining them later in a single "physically irreducible" representation of higher dimension.  As we shall see, this method is unable to capture the full symmetry properties of $m$. In our specific case, this means for example that $m$ can be centrosymmetric even when the inversion operator $I$ does not belong to the magnetic little co-group, as it is always the case if $\textbf{k}$ is a non-Lifshits vector ($2\textbf{k}\notin L^*$ , $L^*$ being the reciprocal lattice). On the other hand, working with $m$ (i.e., the two representations simultaneously) is extremely inconvenient, since, unlike $e^{+i\textbf{k}\cdot\textbf{t}}\psi$, $m$ is not an eigenvector of
the pure translations.   With this in mind, it is useful to reformulate Eq. \ref{pg_condition} as a condition on $e^{+i\textbf{k}\cdot\textbf{t}}\psi$ rather than on $m$:

\begin{equation}
\label{pg_condition_psi} r \in S \longleftrightarrow
\{r|\textbf{v}\} e^{+i\textbf{k}\cdot\textbf{t}}\psi = \pm
\displaystyle\left\{
\begin{array}{c}e^{-i\textbf{k}\cdot\textbf{t}_0}e^{+i\textbf{k}\cdot\textbf{t}}\psi\\or\\e^{+i\textbf{k}\cdot\textbf{t}_0}e^{-i\textbf{k}\cdot\textbf{t}}\psi^*\end{array}\right.
\end{equation}

Eq. \ref{pg_condition_psi} can be further simplified by introducing
the operator of complex conjugation $K$:

\begin{equation}
\label{pg_condition_kappa} r \in S
\longleftrightarrow\displaystyle\left\{\begin{array}{c}\{r|\textbf{v}\}
e^{+i\textbf{k}\cdot\textbf{t}}\psi = \pm
e^{-i\textbf{k}\cdot\textbf{t}_0}e^{+i\textbf{k}\cdot\textbf{t}}\psi\\or\\K\{r|\textbf{v}\}
e^{+i\textbf{k}\cdot\textbf{t}}\psi = \pm
e^{-i\textbf{k}\cdot\textbf{t}_0}e^{+i\textbf{k}\cdot\textbf{t}}\psi\end{array}\right.
\end{equation}

In other words, $e^{+i\textbf{k}\cdot\textbf{t}}\psi$ must be an \emph{eigenvector} of the operator $\{r|\textbf{v}\}$ \emph{or} of the operator $K\{r|\textbf{v}\}$ (or of both), with \emph{eigenvalues} corresponding to $\pm$ Bravais translations. It is worth pointing out already at this stage the crucial difference between \emph{incommensurate} and \emph{commensurate} propagation vectors.  In the former case, \emph{any} eigenvalue will do, since a suitable translation can always be found to equate any phase in the
exponential.  On the contrary, only a finite number of phases are available in the commensurate case. Based on Eq. \ref{pg_condition_kappa}, it becomes natural to consider the mapping not of the group $G$, but of the direct product group $\{E,K\}\otimes G$.  Here, the subtlety is that the images of elements of the form $Kg$ must be antilinear and antiunitary operators \cite{Wigner_1960}.  It is noteworthy that the usefulness of this approach is by no means limited to magnetic structures, and is equally applicable to \emph{structural} modulations, provided that we consider only the "$+$" sign in Eq. \ref{pg_condition} - \ref{pg_condition_kappa}.  Homomorphisms of $\{E,K\}\otimes G$ are known as co-representations (\textit{coreps}) of $G$, and their theory has been extensively developed \cite{Bradley_Cracknell, Kovalev_1993} (see Appendix \ref{Appendix_coreps} for a summary of this theory). In essence, the \textit{corep} analysis consists of 3 steps:
\begin{enumerate}
\item Determination of the the subset $M^\textbf{k}$ of $\{E,K\}\otimes G$ that leaves $\textbf{k}$ invariant,  the equivalent of the "little group" $G^\textbf{k}$.  $M^\textbf{k}$ contains all the operators of the conventional "little group", plus operators of the form $Kg$, where $K$ is the complex conjugation and $g \in G$ exchanges  $\textbf{+k}$ with $\textbf{-k}$.  Consequently $KI \in M^\textbf{k}$ if $I \in G$.
\item Determination of the \textit{coreps} and their image matrices.  The complete analysis has been done by Kovalev \cite{Kovalev_1993} for all space groups and $\textbf{k}$ vectors, and all is required is to refer to the tabulated values therein.
\item Determination of the characteristic (basis) vectors for each \textit{corep}.  This can be done directly, by applying a projection method similar to the standard \textit{irreps}, or, perhaps more easily, by symmetrizing the \textit{irrep} basis vectors, as explained in \cite{Kovalev_1993}.  One must keep in mind that, unlike the case of \textit{irrep} basis vectors, \textit{corep} vectors cannot be multiplied by an arbitrary complex constant, because of the antiunitary character of the associated operators (see below).
\end{enumerate}
Once this analysis is done, the symmetry condition  in Eq. \ref{pg_condition_kappa} can be thoroughly explored by determining the \emph{spectra} of the unitary and antiunitary operators (\emph{images}) associated with the various \textit{coreps}.  Crucially, operators such as the inversion $I$ that exchange $\textbf{+k}$ with $\textbf{-k}$ (not included in the conventional \textit{irrep} analysis) will now be represented by their antiunitary counterparts (e.g., $KI$), which \emph{do} possess an image.  Spectra and eigenvectors for the unitary operators are found in the usual way by diagonalizing the corresponding matrices.  The method to determine the "characteristic vectors" of an antiunitary operator $A$ is described by Wigner \cite{Wigner_1960}.  In particular, it is shown how to construct a full set of orthonormal vectors $v_1 \dots v_n$ that are \emph{invariant} to both $A$ and the unitary operator $A^2$, by linear combinations of the the eigenvectors of $A^2$ with eigenvalue $=+1$.  Linear combinations of the $v_j$'s with \emph{real} coefficients are also invariant by $A$.  Multiplication of $v_j$ (or of a real-coefficient linear combination thereof) by a phase factor $e^{i\omega}$ results in an eigenvector of $A$ with eigenvalue $\lambda=e^{-2i\omega}$.  Linear combinations of eigenvectors with complex coefficients are generally not eigenvectors.  A full spectral analysis of each operator is often not necessary, particularly when the aim is to establish the symmetry of an experimentally determined magnetic structure (see examples below).

\section{Examples}
\label{section: examples}
 In this section, we will analyze the symmetry properties of some magnetic improper ferroelectrics  from the recent literature, using the co-representation approach we described in the previous section. In each case, we will determine the matrix representatives (images) for the relevant propagation vectors/co-representations and the associated basis vectors for the magnetic sites.  We will also determine the point-group structural symmetry for particular ordering patterns.

\subsection{Multiferroic behavior in $RE$MnO$_3$}

The space group is $Pnma$ (no. 62 in the International Tables \cite{IT}; we adopt the standard setting, as opposed to the $Pbnm$ setting used in some papers), and the propagation vector is ($\mu$,0,0), with $\mu$ incommensurate or commensurate but generally in the interior of the Brillouin zone. This propagation vector is labeled as $k_7$ in Kovalev \cite{Kovalev_1993}. We will employ the standard International Tables setting rather than the "Old Kovalev" setting (both are reported in Ref. \cite{Kovalev_1993}).  The small \textit{irreps} for this space group and propagation vectors are all one-dimensional, and their matrices (complex numbers in this case) are reported in Tab. \ref{tab:Pnma_irreps}.

\begin{table*}[h!]
\caption{\label{tab:Pnma_irreps}  Small \textit{irreps} ($\Delta$)
and \textit{coreps} ($D$) of space group $Pnma$ for propagation
vector $k_7=(\mu,0,0)$. The symmetry operators are in the Kovalev
notation and correspond to the International Tables symbols
$h_{1}\equiv 1\hspace{3 pt} 0,0,0$;  $h_2\equiv 2(\frac{1}{2},0,0)\hspace{3 pt} x,\frac{1}{4},\frac{1}{4}$;  $ h_3\equiv 2(0,\frac{1}{2},0)\hspace{3 pt} 0,y,0$;  $h_4\equiv 2(0,0,\frac{1}{2})\hspace{3 pt} \frac{1}{4},0,z$;  $h_{25}\equiv \bar{1}\hspace{3 pt} 0,0,0$; $h_{26}=n(0,\frac{1}{2}, \frac{1}{2})\hspace{3 pt} \frac{1}{4},y,z$; $h_{27}=m\hspace{3 pt} x, \frac{1}{4},z$;  $ h_{28}=a\hspace{3 pt} x,y,\frac{1}{4}$; $\epsilon=e^{+i\pi\mu}$}
\begin{tabular}{ccccc}
& $h_1$&$h_2$& $h_{27}$ & $h_{28}$ \\
& $Kh_{25}$& $Kh_{26}$&$Kh_3$&$Kh_4$\\
\hline
$\Delta_1/D_1$&1& $\epsilon$ & 1 & $\epsilon$\\
$\Delta_2/D_2$&1& $\epsilon$ & -1 & -$\epsilon$\\
$\Delta_3/D_3$&1& -$\epsilon$ & 1 & -$\epsilon$\\
$\Delta_4/D_4$&1& -$\epsilon$ & -1 & $\epsilon$\\
\end{tabular}
\end{table*}

All the \textit{coreps} correspond to "Case a" described in Appendix \ref{Appendix_coreps} ; in other words, each \textit{irrep} generates a single \textit{corep}. In addition, the \textit{coreps} can be set in diagonal form, as explained in Appendix \ref{Appendix_coreps}. The matrices of the antiunitary operators are equal to those of corresponding unitary operators, as shown in Tab. \ref{tab:Pnma_irreps}. The symmetry properties of each \textit{corep} or combination thereof is now clear by inspection of Tab. \ref{tab:Pnma_irreps}, while remembering that the antiunitary operators complex-conjugate all mode coefficients.  In particular:

\begin{enumerate}
\item Linear combination of \textit{corep} modes with purely \emph{real} or purely \emph{imaginary} coefficients are always centric or anti-centric and cannot support ferroelectricity.
\item Multiplication of a centric or anti-centric mode by a phase factor $e^{i\omega}$ is \emph{always} equivalent to a translation for an incommensurate propagation vector but not necessarily so in the
commensurate case. For incommensurate propagation vectors, this analysis confirms that ferroelectricity cannot arise from a \emph{single} magnetic order parameter \cite{Harris_review_2006}. As we shall see in the remainder, however, this is only true in general for 1-dimensional co-representations. \item Linear combination of two \textit{corep} modes with arbitrary \emph{complex} coefficients in general violates all the antiunitary operators and those unitary operators with different matrices for the two coreps. In general this will lead to the polarization vector being allowed in a plane containing the propagation vector.
\item Cycloidal structures are the most important case, because they correspond to the magnetic structures proposed in the literature for the ferroelectric phases. When two components are summed in \emph{quadrature}, they do not always violate all the antiunitary operators. All linear combinations of this kind, with the associated structural point groups and allowed directions of the electrical polarization, are listed in Tab. \ref{tab:cycloidals}.

\begin{table*}[h!]
\caption{\label{tab:cycloidals}  Structural point groups for cycloidal structures of general formula $aD_{\alpha}+ibD_{\beta}$ ($a$ and $b$ are real coefficients). The allowed direction of $\textbf{P}$ is indicated in parenthesis. A dot (.) means that the point group is non-polar, and no ferroelectric polarization can develop.}
\begin{tabular}{ccccc}
& $D_1$&$D_2$& $D_3$ & $D_4$ \\
\hline
$iD_1$&$2mm(x)$& $222(.)$ & $mm2 (z)$ & $m2m (y)$\\
$iD_2$&$222(.)$& $2mm(x)$ & $m2m (y)$ & $mm2(z)$\\
$iD_3$&$mm2 (z)$& $m2m (y)$ & $2mm(x)$ & $222 (.)$\\
$iD_4$&$m2m (y)$&$mm2(z)$ & $222 (.)$ & $2mm(x)$\\
\end{tabular}
\end{table*}

From Tab. \ref{tab:cycloidals}, we can see that the magnetic structure proposed by Kenzelmann \cite{Kenzelmann_2005}, corresponding to the admixture $D_2+iD_3$, only allows a polarization in the $y$ direction ($z$ direction in the $Pbnm$ setting proposed in Ref. \cite{Kenzelmann_2005}), as observed
experimentally. The proposed magnetic structure is therefore consistent with the electrical properties, but no specific magneto-electric mechanism can be inferred from the observation.  It is noteworthy that some combination of \textit{irreps} induce non-centric, non-polar point groups ($222$ in this case).  The
absence of a center of symmetry should not therefore lead to the conclusion that ferroelectricity is allowed in some direction, nor the observation of a cycloidal structure lead to the conclusion that
ferroelectricity is allowed.

\item From this analysis, it is apparent that \emph{commensurate} structures will in general have lower symmetry with respect to corresponding \emph{incommensurate} ones - a well known general result \cite{Toledano}. Here, the obvious reason is that phase factors are not necessarily equivalent to translations. However, the symmetry may be higher for particular choices of the overall phase factor.  An interesting example, which includes the magnetic structure proposed by Aliouane \textit{et al.}, described from the admixture:

\begin{equation}
\psi=e^{i\omega}(aD_1+ibD_3)
\end{equation}

where $a$ and $b$ are real coefficients. This structure is always invariant by application of the mirror plane $\perp b$ ($h_{27}$). Application of the two antiunitary operators $Kh_4$ and $Kh_{26}$ yields (Tab. \ref{tab:Pnma_irreps}):

\begin{equation}
Kh_{26}\psi=Kh_4\psi=\epsilon
e^{-i\omega}(aD_1+ibD_3)=e^{-i(2\omega+\pi \mu)}\psi
\end{equation}

Therefore the corresponding rotations belong to the structural point group only if the phase factor corresponds to $\pm$ a lattice translation, i.e. , if

\begin{equation}
\label{condition_on_omega} \omega=\frac{1}{2}\mu\pi n+2\pi m
\end{equation}

where $n$ and $m$ are arbitrary integers. If $\mu$ is incommensurate, Eq. \ref{condition_on_omega} can always be satisfied to an arbitrary approximation.  In this case, the structural point group is $S=mm2$, and the polarization is along the $z$ axis.  If $\mu$ is commensurate, only a restricted number of phases are available, and Eq. \ref{condition_on_omega} may be far from being satisfied. In this case,  $S=.m.$, and the polarization is in the $x-z$ plane.  The constant-moment "Aliouane" structure corresponds to this case with the particular choices $\mu=\frac{1}{4}$, $a=(2+\sqrt{2})^{\frac{1}{2}}$, $b=(2-\sqrt{2})^{\frac{1}{2}}$ and $\omega=-\frac{\pi}{8}$, for which Eq. \ref{condition_on_omega} is satisfied. Consequently, $S=mm2$ and the polarization must be directed along the $z$ axis. This is also the direction of the polarization found experimentally for the high-field phase of TbMn$O_3$ \cite{Aliouane_2006}.  The microscopic model proposed by Aliouane, based on exchange striction, does in fact produce a $z$ axis polarization, as required by symmetry.

\end{enumerate}

In $RE$MnO$_3$, the Mn atoms are on centers of symmetry, and the application of the antiunitary operators does not generate more sites than those generated by the little group $G^\textbf{k}$. Consequently,  each instance of an \textit{irrep} basis vector generates a single instance of the associated \textit{corep}, spanning exactly the same subspace. Tab. \ref{tab:Pnma_Mn_sites} lists the \textit{corep} modes obtained by symmetrizing the conventional \textit{irrep} modes, following the procedure described in Eq. \ref{transformed_basis}. It can be easily verified that the two modes generated by \ref{transformed_basis} are linearly dependent via a single real coefficient.  The magnetic structure proposed by Kenzelmann \cite{Kenzelmann_2005} corresponds to $m_x(D_3)+im_y(D_2)$, which, as already remarked, only allows the polarization to be along the $y$-axis.  A cycloidal structure of the same type but with spins in the $x-z$ plane would be described as $m_x(D_3)+im_z(D_1)$, which, according to Tab. \ref{tab:cycloidals}, yields a polarization along the $z$-direction ($x$-direction in $Pbnm$). This is consistent with the Ginzburg-Landau analysis performed by Mostovoy \cite{Mostovoy_2006} for the specific case of cycloidal structures.  It is important to remark that the direction of \textbf{P} as established by symmetry does not depend specifically on the direction of the magnetic moments or the type of magnetic structure (cycloidal, helical, etc.)  For example, a helical structure of the type $m_z(D_3)+im_y(D_2)$ and even some complex collinear structures (e.g., $m_y(D_3)+im_y(D_2)$) have exactly the same symmetry as the Kenzelmann \cite{Kenzelmann_2005} structure. Naturally, magnetic measurements and neutron diffraction can all be used to distinguish between these possibilities, and to guide the analysis towards a microscopic model.

\begin{table*}[h!]
\caption{\label{tab:Pnma_Mn_sites} Magnetic (axial vector)\textit{corep} modes for the perovskite $\textit{B}$-site ($Mn$ in the case of $RE$MnO$_3$) associated with the four type-a irreducible co-representations for space group $Pnma$ and propagation vector $k_7=(\mu,0,0)$.  Note the similarity of these modes with those listed in ref. \cite{Brinks_2001}, (Tab. III),  and references cited therein. However, through \textit{corep} analysis, we have \emph{specifically} enforced invariance by application of the antiunitary operator $KI$, where $I$ is the inversion at the origin of the coordinate system and $K$ is the complex conjugation.  This invariance can only be accidental in \textit{irrep} analysis.  The matrix elements for unitary and antiunitary operators can be found in Tab. \ref{tab:Pnma_irreps}.  $\epsilon=e^{+i\pi\mu}$ and $\epsilon^*$ is its complex conjugate.}
\begin{ruledtabular}
\begin{tabular}{cccccc}
& &$Mn(1)=\frac{1}{2},0,0$& $Mn(2)=0,\frac{1}{2},\frac{1}{2}$&$Mn(3)=\frac{1}{2},\frac{1}{2},0$&$Mn(4)=0,0,\frac{1}{2}$ \\
\hline
$D_1$&$m_x$ &$\epsilon^*$&1 & -$\epsilon^*$ & -1\\
&$m_y$&$\epsilon^*$&-1 & $\epsilon^*$ & -1\\
&$m_z$&$\epsilon^*$&-1 & -$\epsilon^*$ & 1\\
\hline
$D_2$&$m_x$ &$\epsilon^*$&1 & $\epsilon^*$ & 1\\
&$m_y$&$\epsilon^*$&-1 & -$\epsilon^*$ & 1\\
&$m_z$&$\epsilon^*$&-1 & $\epsilon^*$ & -1\\
\hline
$D_3$&$m_x$ &$\epsilon^*$&-1 & -$\epsilon^*$ & 1\\
&$m_y$&$\epsilon^*$&1 & $\epsilon^*$ & 1\\
&$m_z$&$\epsilon^*$&1 & -$\epsilon^*$ & -1\\
\hline
$D_4$&$m_x$ &$\epsilon^*$&-1 & $\epsilon^*$ & -1\\
&$m_y$&$\epsilon^*$&1 & -$\epsilon^*$ & -1\\
&$m_z$&$\epsilon^*$&1 & $\epsilon^*$ & 1\\
\end{tabular}
\end{ruledtabular}
\end{table*}

The case of the $RE$ sites is more interesting, because the atoms do not sit on a center of symmetry, and they are therefore split into orbits by the little group $G^\textbf{k}$. The application of the antiunitary operators mixes the two orbits, so that the \textit{corep} modes are combination of \textit{irrep} modes on the two orbits.  In this case, the two modes generated by \ref{transformed_basis} are linearly independent. Tab. \ref{tab:Pnma_RE_sites} lists the \textit{corep} modes obtained by symmetrizing the conventional \textit{irrep} modes, following the procedure described in Eq. \ref{transformed_basis}.

\begin{table*}[h!]
\caption{\label{tab:Pnma_RE_sites} Magnetic (axial vector) \textit{corep} modes for the perovskite $\textit{A}$-site ($RE$ in the case of $RE$MnO$_3$) associated with the four type-a irreducible co-representations for space group $Pnma$ and propagation vector $k_7=(\mu,0,0)$.  By construction, these modes are invariant by application of the antiunitary operator $KI$, where $I$ is the inversion at the origin of the coordinate system and $K$ is the complex conjugation.  The matrix elements for unitary and antiunitary operators can be found in Tab. \ref{tab:Pnma_irreps}.$\epsilon=e^{+i\pi\mu}$ and $\epsilon^*$ is its complex conjugate.}
\begin{ruledtabular}
\begin{tabular}{cccccc}
& &$RE(1)=x,\frac{1}{4},z$& $RE(2)=x+\frac{1}{2},\frac{1}{4},-z+\frac{1}{2}$&$RE(3)=-x+\frac{1}{2},\frac{3}{4},z-\frac{1}{2}$&$RE(4)=-x+1,-\frac{1}{4},-z+1$ \\
\hline
$D_1$&$m_y$ &1&$\epsilon^*$ & 1 & $\epsilon^*$\\
&$m_y'$&1&$\epsilon^*$ & -$i$ & -$i\epsilon^*$\\
\hline
$D_2$&$m_x$ &1&-$\epsilon^*$ & -1 & $\epsilon^*$\\
&$m_x'$&1&-$\epsilon^*$ & $i$ & -$i\epsilon^*$\\
&$m_z$ &1&$\epsilon^*$ & 1 & $\epsilon^*$\\
&$m_z'$&1&$\epsilon^*$ & -$i$ & -$i\epsilon^*$\\
\hline
$D_3$&$m_x$ &1&$\epsilon^*$ & 1 & $\epsilon^*$\\
&$m_x'$&1&$\epsilon^*$ & -$i$ & -$i\epsilon^*$\\
&$m_z$ &1&-$\epsilon^*$ & -1 & $\epsilon^*$\\
&$m_z'$&1&-$\epsilon^*$ & $i$ & -$i\epsilon^*$\\
\hline
$D_4$&$m_y$ &1&-$\epsilon^*$ & -1 & $\epsilon^*$\\
&$m_y'$&1&-$\epsilon^*$ & $i$ & -$i\epsilon^*$\\
\end{tabular}
\end{ruledtabular}
\end{table*}

\subsection{BiMn$_2$O$_5$ and DyMn$_2$O$_5$}

The space group is $Pbam$ (no. 55 in the International Tables \cite{IT}), with three relevant Wyckoff sites:  4f $(0,\frac{1}{2},z)$ with $z\approx\frac{1}{4}$  for the $Mn^{4+}$ sites, 4h $(x,y,\frac{1}{2})$ for the $Mn^{3+}$ and 4g $(x,y,0)$ for the $RE$ sites.  4h and 4g have the same symmetry, and can be
treated in a completely analogous way. The propagation vector is ($\frac{1}{2}$,0,0), $k_{20}$ in Kovalev notation, for the low-temperature phase of $Dy$ \cite{Wilkinson_DyMn2O5} (the minority component ($\frac{1}{2}$,0,$\mu$) will be dealt with in the next section) and ($\frac{1}{2}$,0,$\frac{1}{2}$), $k_{24}$ for $Bi$ \cite{Munoz_2002}.  Both propagation vectors are special points of the Brillouin zone for which $k \equiv -k$, and have identical irrep/coreps. Moreover, as explained in Section
\ref{section:theory}, the inversion operator $I \in G_\textbf{k}$, and there is no need to introduce antiunitary operators.  The analysis is therefore very similar to the one performed by Munoz
\cite{Munoz_2002} for the Bi case.  It is convenient to perform a unitary transformation of the Kovalev matrix representatives and associated modes, through the unitary matrix

\begin{equation}
\label{unitary_transf} U=\frac{1+i}{2}\left( \begin{array}{cc}  i     &  1      \\
1 & i
\end{array}  \right)
\end{equation}

With this transformation, the resulting matrices (Tab. \ref{Tab:Pbam_comm_irrep}) become \emph{real}. For site 4f, the modes (Tab. \ref{Tab:Pbam_comm_modes_4f}) can also be made real by multiplication of each subspace basis set by an appropriate coefficient.  For site 4h, a bit more care is required, since each of the two irreps appears twice for every spin direction, so there is a degree of arbitrariness in the definition of the invariant subspaces.  Here, we have chosen the definition so that the basis vectors have constant moments on all sites, but other choices are possible.  Physically, this means that the magnetic moments are related in pairs (S1 is related to S2 and S3 to S4), but the pairs are allowed in principle to have different moments within the same irrep.

\begin{table*}[h!]
 \caption{\label{Tab:Pbam_comm_irrep}Matrix representatives of the irreducible representations  of the little group G$^k$ for the space group $G=Pbam$ and $k_{20}=(\frac{1}{2},0,0)$, or $k_{24}=(\frac{1}{2}$,0,$\frac{1}{2})$. The matrices reported herein are the same as in Ref. \cite{Munoz_2002}, and are related to the Kovalev matrices by the unitary transformation $UMU^{-1}$, where $U$ is given in Eq. \ref{unitary_transf}.}
\begin{ruledtabular}
\begin{tabular}[b]{c|*{8}{c}}
 Irreps & h$_1$ & h$_2$ & h$_{3}$ & h$_{4}$ & h$_{25}$ & h$_{26}$ & h$_{27}$ & h$_{28}$ \\
 \hline
 $\Gamma_1$ &
 $ \left( \begin{array}{cc}  1     &  0      \\   0     &  1      \end{array}  \right) $ &
 $ \left( \begin{array}{cc}  0     &  1      \\   -1     &  0      \end{array}  \right) $ &
 $ \left( \begin{array}{cc}   1     &  0      \\   0     &   -1    \end{array}  \right) $ &
 $ \left( \begin{array}{cc}  0     &  1      \\   1     &  0      \end{array}  \right) $ &
 $ \left( \begin{array}{cc}  0     &  1     \\   1     &  0      \end{array}  \right) $ &
 $ \left( \begin{array}{cc}  1     &  0      \\   0     &  -1      \end{array}  \right) $ &
 $ \left( \begin{array}{cc}  0     &  1      \\   -1     &  0      \end{array}  \right) $ &
 $ \left( \begin{array}{cc}  1     &   0      \\  0     &  1      \end{array}  \right) $ \\
  $\Gamma_2$ &
$ \left( \begin{array}{cc}  1     &  0      \\   0     &  1\end{array}  \right) $ &
 $ \left( \begin{array}{cc}  0     &  1      \\   -1     &  0      \end{array}  \right) $ &
 $ \left( \begin{array}{cc}   1     &  0      \\   0     &   -1    \end{array}  \right) $ &
 $ \left( \begin{array}{cc}  0     &  1      \\   1     &  0      \end{array}  \right) $ &
 $ \left( \begin{array}{cc}  0     &  -1     \\   -1     &  0      \end{array}  \right) $ &
 $ \left( \begin{array}{cc}  -1     &  0      \\   0     &  1      \end{array}  \right) $ &
 $ \left( \begin{array}{cc}  0     &  -1      \\   1     &  0      \end{array}  \right) $ &
 $ \left( \begin{array}{cc}  -1     &   0      \\  0     &  -1      \end{array}  \right) $ \\
\end{tabular}
\end{ruledtabular}
\end{table*}

The key aspect to assess the symmetry of the possible magnetic structures is the fact that both \textit{irreps} are \emph{2-dimensional}.  Therefore, in contrast with the REMn$O_3$ example, it is possible to obtain non-centrosymmetric and polar structures even for a \emph{single} order parameter, provided that certain special directions in the 2-dimensional space are avoided. The magnetic basis vectors for the propagation vector $k_{24}=(\frac{1}{2}$,0,$\frac{1}{2})$ are reported in Tab. \ref{Tab:Pbam_comm_modes_4f} and Tab. \ref{Tab:Pbam_comm_modes_4h} for the two sites.  Modes for $k_{20}=(\frac{1}{2},0,0)$ are slightly different using our atom conventions, but can be obtained
in the same straightforward way (see Table captions).

\begin{table*}[h!]
\caption{\label{Tab:Pbam_comm_modes_4f} Magnetic (axial vector) \textit{irrep} modes for the 4f $(0,\frac{1}{2},z)$ sites ($Mn^{4+}$ in $RE$Mn$_2$O$_5$) of space group $G=Pbam$ and $k_{24}=(\frac{1}{2}$,0,$\frac{1}{2})$.  The basis vectors reported herein are related to the Kovalev basis vectors (obtained by the projection method with the help of the program Sarah \cite{Sarah} by the unitary transformation $vU^{-1}$, where $U$ is given in Eq. \ref{unitary_transf}. The matrix elements can be found in Tab. \ref{Tab:Pbam_comm_irrep}.  Primed and unprimed modes (e.g., $m_x$ and $m'_x$ belong to the same invariant subspace, and have been both multiplied by the same coefficient, so that the unprimed mode is always 1 on atom 1.}
\begin{ruledtabular}
\begin{tabular}{cccccc}
& &$Mn^{4+}(1)=0,\frac{1}{2},\frac{1}{4}$& $Mn^{4+}(2)=\frac{1}{2},0,\frac{3}{4}$&$Mn^{4+}(3)=0,\frac{1}{2},\frac{3}{4}$&$Mn^{4+}(4)=\frac{1}{2},0,\frac{1}{4}$ \\
\hline
$\Gamma_1$&$m_x$ &1&-1 & -1 & 1\\
&$m_x'$&-1&-1 & 1 & 1\\
&$m_y$&1&1 & -1 & -1\\
&$m_y'$&-1&1 & 1 & -1\\
&$m_z$&1&-1 & 1 & -1\\
&$m_z'$&1&1 & 1 & 1\\
\hline
$\Gamma_2$&$m_x$ &1&-1 & 1 & -1\\
&$m_x'$&-1&-1 & -1 & -1\\
&$m_y$&1&1 & 1 & 1\\
&$m_y'$&-1&1 & -1 & 1\\
&$m_z$&1&-1 & -1 & 1\\
&$m_z'$&1&1 & -1 & -1\\

\end{tabular}
\end{ruledtabular}
\end{table*}

\begin{table*}[h!]
\caption{\label{Tab:Pbam_comm_modes_4h} Magnetic (axial vector) \textit{irrep} modes for the 4h $(x,y,\frac{1}{2})$ $Mn^{3+}$ sites ($x \sim 0.1$, $y \sim 0.85$) of space group $G=Pbam$ and $k_{24}=(\frac{1}{2}$,0,$\frac{1}{2})$. The $RE$ atoms are on site 4g with the same site symmetry, and their modes can be deduced in a completely analogous way. The basis vectors reported herein are related to the Kovalev basis vectors (obtained by the projection method with the help of the program Sarah \cite{Sarah} by the unitary transformation $vU^{-1}$, where $U$ is given in Eq. \ref{unitary_transf}. The Sarah modes were preliminary recombined across invariant subspaces so as to have constant moments on all sites.  The matrix elements can be found in Tab. \ref{Tab:Pbam_comm_irrep}. Primed and unprimed modes (e.g., $m1_x$ and $m1'_x$) belong to the same invariant subspace, and have been both multiplied by the same coefficient, so that the unprimed mode is always 1 on atom 1.  }
\begin{ruledtabular}
\begin{tabular}{cccccc}
& &$Mn^{3+}(1)=x,y,\frac{1}{2}$& $Mn^{3+}(2)=1-x,y-\frac{1}{2},\frac{1}{2}$&$Mn^{3+}(3)=\frac{1}{2}-x,y-\frac{1}{2},\frac{1}{2}$&$Mn^{3+}(4)=\frac{1}{2}+x,\frac{3}{2}-y,\frac{1}{2}$ \\
\hline
$\Gamma_1$&$m1_x$ &1&1 & 1 & -1\\
&$m1_x'$&1&-1 & -1 & -1\\
&$m2_x$&1&1 & -1 & 1\\
&$m2_x'$&-1&1 & -1 & -1\\
&$m1_y$&1&-1 & -1 & -1\\
&$m1_y'$&1&1 & 1 & -1\\
&$m2_y$&1&1 & -1 & 1\\
&$m2_y'$&-1&1 & -1 & -1\\
\hline
$\Gamma_2$&$m1_z$ &1&-1 & 1 & 1\\
&$m1_z'$&1&1 & -1 & 1\\
&$m2_z$&1&1 & 1 & -1\\
&$m2_z'$&-1&1 & 1 & 1\\

\end{tabular}
\end{ruledtabular}
\end{table*}

The experimentally determined magnetic structure belongs to the $\Gamma_1$  \textit{irrep}, and is a combination of $m_x$ and $m_y$ modes for the 4f sites and $m1_x$ and $m2_y$ for the 4h sites.  The important point is that only one \emph{unprimed} basis vectors of each subspace is employed in each instance, so that all the components transform in the same way:

\begin{eqnarray}\label{exp_modes_Bimn2o5} \psi(Mn^{4+})&=&c_1m_x+c_2m_y'=c_1 \left[m_x, m_x'\right]\cdot \left[ \begin{array}{c}  1   \\
0
\end{array}  \right]+c_2 \left[m_y, m_y'\right]\cdot \left[ \begin{array}{c}  1   \\
0
\end{array}  \right]\\
\psi(Mn^{3+})&=& c_3m1_x+c_4m2_y= c_3\left[m1_x, m1_x'\right]\cdot \left[ \begin{array}{c}  1   \\
0
\end{array}  \right]+c_4\left[m2_y, m2_y'\right]\cdot \left[ \begin{array}{c}  1   \\
0
\end{array}  \right]\nonumber
\end{eqnarray}

The matrices are applied precisely to the column vectors in Eq. \ref{exp_modes_Bimn2o5}.  With this notation, the symmetry can be read straightforwardly from the \emph{irrep} matrices in Tab. \ref{Tab:Pbam_comm_irrep}.  It is clear that only diagonal matrices ($h_3$, $h_{26}$, $h_{28}$) survive, because off-diagonal matrices transform unprimed into primed modes.  The structural point-group symmetry in the magnetically ordered phase is therefore $m2m$, which allow the polarization only along the $b$ axis, as found experimentally.

\subsection{TbMn$_2$O$_5$ - commensurate phase}

TbMn$_2$O$_5$ is isostructural to the previous compounds, but orders magnetically with different propagation vectors \cite{Chapon1, Blake}. Here, we will only consider the \emph{commensurate},
high-temperature phase, with ($\frac{1}{2}$,0,$\frac{1}{4}$), $k_{16}$ in Kovalev notation, but the same analysis would apply to an incommensurate phase of the type ($\frac{1}{2}$,0,$\mu$), also labeled $k_{16}$. There are only 3 elements in the \textit{irrep} little group: a two-fold axis $\parallel$ to $z$ ($h_4$) and two glide planes $\perp$ $x$ ($h_{26}$) and $y$ ($h_{27}$). There is a single 2-dimensional irrep. The $\beta$ matrix is the identity, so the matrices for the antiunitary operators $Kh_{25}$, $Kh_{2}$, $Kh_{3}$ and $Kh_{28}$ are the same as those for $h_{1}$ (the identity), $h_{26}$, $h_{27}$ and $h_{4}$, respectively.  The basis vectors can be constructed using Eq. \ref{transformed_basis}, and are obviously invariant by inversion. The 4f $(0,\frac{1}{2},\frac{1}{4})$ ($Mn^{4+}$) sites are split by the
irreps into two orbits, which are recombined to obtain the basis functions for the \textit{coreps}, whereas site 4h $(x,y,\frac{1}{2})$ remains as a single orbit and Eq. \ref{transformed_basis} produces sets of degenerate vectors.  At this stage, it is useful to perform a unitary transformation using the matrix

\begin{equation}
\label{unitary_transf_k16} U=\left( \begin{array}{cc}  1     &  0      \\
0 & i
\end{array}  \right)
\end{equation}

which makes all the matrices real.  Note the special form of the unitary transformations for antiunitary operators (Eq. \ref{similarity}).  The resulting \textit{corep} matrices are reported in Tab. \ref{Tab:Pbam_comm_irrep_k16}.

\begin{table*}[h!]
\caption{\label{Tab:Pbam_comm_irrep_k16} Matrix representatives of the irreducible co-representation for the space group $G=Pbam$ and $k_{16}=(\frac{1}{2},0,\mu)$ ($\mu=\frac{1}{4}$ for the commensurate phase of $TbMn_2O_5$). The Kovalev matrices (the same for pairs of unitary and antiunitary operators, as explained in the text) were transformed using the unitary matrix  from Eq. \ref{unitary_transf_k16}.}
\begin{ruledtabular}
\begin{tabular}[b]{c|*{8}{c}}
 Coreps & h$_1$ & h$_4$ & h$_{26}$ & h$_{27}$ & Kh$_{25}$ & Kh$_{2}$ & Kh$_{3}$ & Kh$_{28}$ \\
 \hline
 $D_1$&$ \left( \begin{array}{cc}  1     &  0      \\   0     &  1      \end{array}  \right) $ &
 $ \left( \begin{array}{cc}  1     &  0      \\   0     &  -1      \end{array}  \right) $ &
 $ \left( \begin{array}{cc}   0     &  1      \\   1     &   0    \end{array}  \right)
 $ &
 $ \left( \begin{array}{cc}  0     &  -1      \\   1     &  0     \end{array}  \right) $ &
 $ \left( \begin{array}{cc}  1     &  0     \\   0     &  -1      \end{array}  \right) $ &
 $ \left( \begin{array}{cc}  0     &  1      \\   -1     &  0      \end{array}  \right) $ &
 $ \left( \begin{array}{cc}  0     &  -1      \\   -1     &  0      \end{array}  \right) $ &
 $ \left( \begin{array}{cc}  1     &   0      \\  0     &  1      \end{array}  \right) $ \\
\end{tabular}
\end{ruledtabular}
\end{table*}

The basis vectors can be derived in the same way, by applying the inverse unitary transformation to the \textit{corep} basis from Eq. \ref{transformed_basis}, and are reported in Tab. \ref{Tab:Pbam_comm_modes_4f_k16} and \ref{Tab:Pbam_comm_modes_4h_k16}.

\begin{table}
\caption{\label{Tab:Pbam_comm_modes_4f_k16} Magnetic (axial vector) \textit{corep} modes for the 4f $(0,\frac{1}{2},z)$ sites ($Mn^{4+}$ in $REMn_2O_5$) of space group $G=Pbam$ and $k_{16}=(\frac{1}{2}$,0,$\frac{1}{4})$.  Primed and unprimed modes (e.g., $m1_x$ and $m1'_x$ belong to the same invariant subspace.}
\begin{ruledtabular}
\begin{tabular}{cccccc}
&$Mn^{4+}(1)=0,\frac{1}{2},\frac{1}{4}$& $Mn^{4+}(2)=\frac{1}{2},0,\frac{3}{4}$&$Mn^{4+}(3)=0,\frac{1}{2},\frac{3}{4}$&$Mn^{4+}(4)=\frac{1}{2},0,\frac{1}{4}$ \\
\hline
$m1_x$ &0& -1 & 0 & i\\
$m1_x'$&i&0 & -1 & 0\\
$m2_x$&0&i & 0 & -1\\
$m2_x'$&-1&0 & i & 0\\
$m1_y$&0&-1 &0 & i\\
$m1_y'$&-i&0 & 1 & 0\\
$m2_y$ &0&i & 0 & -1\\
$m2_y'$&1&0 & -i & 0\\
$m1_z$&1&0 & i & 0\\
$m1_z'$&0&-i & 0 & -1\\
$m2_z$&i&0 & 1 & 0\\
$m2_z'$&0&-1 & 0 & -i\\

\end{tabular}
\end{ruledtabular}
\end{table}

\begin{table}
\caption{\label{Tab:Pbam_comm_modes_4h_k16} Magnetic (axial vector) \textit{correp} modes for the 4h $(x,y,\frac{1}{2})$ $Mn^{3+}$ sites ($x \sim 0.1$, $y \sim 0.85$) of space group $G=Pbam$ and $k_{16}=(\frac{1}{2}$,0,$\frac{1}{4})$. The $RE$ atoms are on site 4g with the same site symmetry, and their modes can be deduced in a completely analogous way. Primed and unprimed modes (e.g., $m1_x$ and $m1'_x$) belong to the same invariant subspace. $\gamma=\displaystyle \frac{1+i}{\sqrt{2}}$, and $\gamma^*$ is its complex conjugate.}
\begin{ruledtabular}
\begin{tabular}{cccccc}
&$Mn^{3+}(1)=x,y,\frac{1}{2}$& $Mn^{3+}(2)=1-x,y-\frac{1}{2},\frac{1}{2}$&$Mn^{3+}(3)=\frac{1}{2}-x,y-\frac{1}{2},\frac{1}{2}$&$Mn^{3+}(4)=\frac{1}{2}+x,\frac{3}{2}-y,\frac{1}{2}$\\
\hline
$m1_x$ &$\gamma^*$&$\gamma^*$ & 0 & 0\\
$m1_x'$&0&0 & $\gamma^*$ & -$\gamma^*$\\
$m2_x$&0&0 & $\gamma^*$ & $\gamma^*$\\
$m2_x'$&$\gamma^*$&-$\gamma^*$ & 0 & 0\\
$m1_y$&$\gamma^*$&$\gamma^*$ & 0 & 0\\
$m1_y'$&0&0 & -$\gamma^*$ & $\gamma^*$\\
$m2_y$&0&0 & $\gamma^*$ & $\gamma^*$\\
$m2_y'$&-$\gamma^*$&$\gamma^*$ & 0 & 0\\
$m1_z$ &0&0 & $\gamma$ & -$\gamma$\\
$m1_z'$&-$\gamma$&-$\gamma$ & 0 & 0\\
$m2_z$&$\gamma$&-$\gamma$ & 0 & 0\\
$m2_z'$&0&0 & -$\gamma$ & -$\gamma$\\

\end{tabular}
\end{ruledtabular}
\end{table}

It is easy to show that the constant-moments experimental solution \cite{Chapon1, Blake, Chapon2} involves mixing modes spanning the same subspace with equal or opposite coefficients.  Only the $m1_x$/$m1_x'$ and to a lesser extent the $m1_y$/$m1_y'$ components are relevant on both sites, so that, for example,  the x-components $\psi_x$ of sites $Mn^{4+}$ and $Mn^{3+}$ are:

\begin{eqnarray}\label{exp_modes_tbmn2o5} \psi_x(Mn^{4+})&=&\gamma^* (m1_x+m1_x')=\gamma^* \left[m1_x, m1_x'\right]\cdot \left[ \begin{array}{c}  1   \\
1
\end{array}  \right]\\
\psi_x(Mn^{3+})&=& (m1_x+m1_x')= \left[m1_x, m1_x'\right]\cdot \left[ \begin{array}{c}  1   \\
1
\end{array}  \right]\nonumber
\end{eqnarray}

where $\gamma=\displaystyle \frac{1+i}{\sqrt{2}}$.  Once again, the matrices are applied precisely to the column vectors in Eq. \ref{exp_modes_tbmn2o5}.  With this notation, the symmetry can be read off directly from Tab \ref{Tab:Pbam_comm_irrep_k16}.  Note that the phase factor $\gamma^*$ in Eq. \ref{exp_modes_tbmn2o5} does not reduce the symmetry even in the commensurate case, because $\gamma^2$ is a lattice translation. The surviving operators h26, Kh3 and Kh28 define the structural point group symmetry $m2m$ ($C_{2v}$), indicating that $b$ is the only polar direction, as observed experimentally.  Note that the conclusion is identical to the BiMn$_2$O$_5$ case, in spite of the fact that we have adopted different basis conventions for the invariant subspaces.  This is a direct confirmation of our conjecture \cite{Blake} that the $c$-axis component of the propagation vector does not affect the symmetry properties of the system, provided that the in-plane magnetic structure is the same.

\subsection{HgCr$_2$S$_4$}

Recently, attention has been drawn to chalcogenide chromium spinels (cubic, space group $Fd\bar{3}m$) of the type $A$Cr$_2$X$_4$ ($A$=Cd,Hg; X=S,Se) which are weakly ferroelectric in their magnetically ordered state and have been classify as multiferroic materials\cite{CdCr2S4Nature,PhysRevLett.96.157202,PhysRevB.73.224442}. HgCr$_2$S$_4$ is particularly interesting, because it has a complex magnetic structure, whereas all the other chalcogenide spinels are ferromagnetic.  Very recently \cite{Chapon3}, we have studied the HgCr$_2$S$_4$ magnetic structure using high-resolution neutron powder diffraction. Long-range incommensurate magnetic order sets in at T$_N\sim$22K with propagation vector \textbf{k}=(0,0,$\sim$0.18). On cooling below T$_N$, the propagation vector increases and saturates at the commensurate value \textbf{k}=(0,0,0.25). The magnetic structure below T$_N$ consists of ferromagnetic layers in the \textit{ab}-plane stacked in a helical arrangement along the \textit{c}-axis.  We also performed a full symmetry analysis using co-representations, determining the matrices and modes for the relevant \textit{corep}, which is derived from the $\Gamma_5$ \textit{irrep} as explained above.  We will not repeat the detailed analysis, referring instead to our previous paper \cite{Chapon3}, of which we summarize the salient point herein. There are 4 \textit{corep} basis vectors, corresponding to pairs of ferromagnetic and antiferromagnetic helices, all with the moments in the $x-y$ plane.  Pairs of basis vectors defining invariant subspaces correspond to left- and right-handed helices. The experimental solution is one of the ferromagnetic helices:  as in the previous case, the symmetry for an incommensurate propagation vector can be read off the matrices in Ref. \cite{Chapon3}, Tab. II, considering that only those in diagonal form ($h_1$, $h_4$, $h_{14}$, $h_{15}$, $Kh_2$, $Kh_3$,$Kh_{13}$ and $Kh_{16}$) survive below the ordering temperature. These operators define the structural point group $422$, which is \emph{non-polar}, in spite of being non-centrosymmetric. We therefore conclude that ferroelectricity cannot arise from the magnetic transition at T$_N\sim$22K to an incommensurate phase, although the observed polarization can very well have other causes. The low-temperature phase is close to be perfectly commensurate with \textbf{k}=(0,0,0.25), and the assessment of its symmetry requires additional care, since, as already illustrated, it depends on the overall phase factor of the magnetic structure. As a preliminary observation, we remark that the lattice is \textit{F}-centered, so lattice translations have eigenvalues that are multiple of $e^{i\frac{\pi}{4}}$. The 4-fold rotations $h_{14}$ and $h_{15}$ are always lost, because $i\epsilon=e^{i\frac{3}{8}\pi}$ is not a lattice translation.  The symmetry is always lowest for a generic phase factor, for which only $h_4$ survives (point group $2$). However, pairs of orthogonal 2-fold axes survive for a global phase of $0$ mod
$\frac{\pi}{8}$ ($Kh_{13}$ and $Kh_{16}$) or $\frac{\pi}{16}$ mod $\frac{\pi}{8}$ ($Kh_{2}$ and $Kh_{3}$), so that the point-group symmetry is $222$ in both cases.

\section{Discussion and conclusions}
\label{section: discussion}

\indent We have presented a general method, based on a non-conventional application of co-representation analysis, to determine the point-group symmetry below a magnetic ordering transition for a crystal that is centrosymmetric in the paramagnetic phase, regardless of the direction and magnitude of the magnetic propagation vector. This method was employed to determine the constraints on the development of electrical polarization in a new class of "magnetic improper" multiferroics. This approach can be readily extended, with essentially no modifications, to paramagnetic crystals with non-polar, non-centrosymmetric groups. It is also noteworthy that the point group we derive can be employed to set constraints on other vector or tensor quantities in the magnetically ordered phase. Pragmatically, we found the process of deriving \textit{corep} matrices to be straightforward, when one has become familiar with the Kovalev tables \cite{Kovalev_1993}. The only \textit{caveat} is that one should be careful in dealing with Eq. \ref{a_type_transform}, since the operator $a_0ga_0^{-1}$ may be related to $g$ by a translation, entailing an additional phase factor in the matrices.  We found that the most effective way of deriving \textit{corep} basis vectors is to transform the \textit{irrep} basis vectors as in Eq. \ref{general_transformed_basis}. The \textit{irrep} basis vectors can be obtained directly by projection, or with the assistance of one of several dedicated programs such as FullProF, Mody or Sarah \cite{Sarah,mody,fulprof}. Whatever the method employed, it is useful to check the symmetry of the basis vectors against the \textit{corep} matrices.  This is best done graphically, by superimposing the basis vector pattern, with the appropriate phase factors indicated, onto a diagram of the symmetry elements as in the International Tables \cite{IT}.
With a savvy choice of the basis vectors, the structural point-group symmetry of the magnetically ordered phase can often be deduced by inspection for a variety of physically relevant magnetic structures\cite{IT}.

\bibliography{PGR_symmetry}

\appendix

\section{An overview of co-representation theory}
\label{Appendix_coreps}

\textit{Coreps} are constructed in a very similar way to
\textit{irreps}, on the basis of the subset $M^\textbf{k}$ of
$\{E,K\}\otimes G$ that leaves $\textbf{k}$ invariant,  the
equivalent of the "little group" $G^\textbf{k}$.  It is important to
stress that, by its very construction, $M^\textbf{k}$ (like
$G^\textbf{k}$) never mixes the $+\textbf{k}$ and $-\textbf{k}$
Fourier components, so the basis functions obtained through
\textit{corep} analysis are trivial eigenvalues of the pure
translations. We can therefore drop the prefix
$e^{i\textbf{k}\cdot\textbf{t}}$ in Eq. \ref{pg_condition_kappa} and
limit our analysis to the \emph{representative} elements of
$M^\textbf{k}$ as for usual \textit{irreps}. This accomplishes our
goal of capturing the full symmetry properties of $m$ whilst
conveniently working with Fourier components. We can distinguish
three cases:

\begin{enumerate}
\item $M^\textbf{k}= G^\textbf{k}$.  This occurs when
$-\textbf{k}$ is not in the \textit{irrep} star of $\textbf{k}$,
which is the case only for certain non-centrosymmetric space groups.
By their very definition, the magnetic improper multiferroic
materials we deal with here are always centrosymmetric in the
paramagnetic phase, and this case is therefore never relevant. \item
$M^\textbf{k}= G^\textbf{k}+KG^\textbf{k}$.  Since
$K\textbf{k}=-\textbf{k}$ and $KE \in M^\textbf{k}$,  $\textbf{k}$
and $-\textbf{k}$ must be equivalent, i.e., $\textbf{k}$ must be a
Lifshits vector. The use of \textit{coreps} resolves the difficulty
with inequivalent complex conjugate representations, which in the
conventional \textit{irrep} analysis are somewhat artificially
combined in "physical" \textit{irreps}. However, \textit{coreps} do
not add much in terms of symmetry analysis, since all the relevant
proper and improper rotations are already contained in
$G^\textbf{k}$. \item $M^\textbf{k}= G^\textbf{k}+a_0G^\textbf{k}$,
where $a_0=Kh$ and $h \in (G-G^\textbf{k})$.  Once again, the very
nature of our problem dictates  that the inversion $I \in G$, so we
can always choose $a_0=KI$. In this case, it is clear that
$M^\textbf{k}$ contains more rotations than $G^\textbf{k}$, and
\textit{coreps} should always be used.
\end{enumerate}

The small \textit{coreps} of $M^\textbf{k}$ are built out of pairs
of small \textit{irreps} of $G^\textbf{k}$, $\Delta(g)$ and
$\overline{\Delta}(g)={\Delta(a_0^{-1}ga_0)}^*$ and their
corresponding basis vectors $\psi$ and $\phi=a_0\psi$.  Note that
$\psi$ and $\phi$ may be linearly dependent, but are always
independent if the magnetic atom is split into orbits or if
$\Delta(g)$ and $\overline{\Delta}(g)$ are not equivalent.  Using
the compound basis $\langle \psi, \phi|$, the derived \textit{corep}
is always of diagonal form for $g \in G^\textbf{k}$ and of
off-diagonal form for $a_0g \in (M^\textbf{k}-G^\textbf{k})$:

\begin{eqnarray}
\label{primitive_matrix}
D(g)&=&\left[\begin{array}{cc}\Delta(g)&0\\0&\overline{\Delta}(g)\end{array}\right]\\
D(a_0g)&=&\left[\begin{array}{cc}0&\Delta(a_0ga_0)\\\Delta(g)^*&0\end{array}\right]\nonumber
\end{eqnarray}

The matrices in Eq. \ref{primitive_matrix} may be reducible to a
simpler form by a change of basis through a unitary operator $U$.
Note that the similarity transformation has a different form for
antiunitary operators:

\begin{eqnarray}
\label{similarity}
D'(g)&=UD(g)U^{-1} \hspace{6 pt} &\forall g \in G^\textbf{k}\\
D'(a_0g)&=U^*D(a_0g)U^{-1} \hspace{6 pt} &\forall a_0g \in
(M^\textbf{k}-G^\textbf{k})\nonumber
\end{eqnarray}

We can distinguish three further cases:

\begin{description}
\item[a \hspace{6 pt}] $\Delta(g)$ and $\overline{\Delta}(g)$ are equivalent through
a unitary matrix $N$, so that
$\Delta(g)=N\overline{\Delta}(g)N^{-1}$ \emph{and}
$NN^*=+\Delta(a_0^2)$.  In this case, Eq. \ref{primitive_matrix} can
be reduced to block diagonal form for both unitary and antiunitary
operators.  A significant simplification occurs if $N$ is the
identity matrix.  This is always the case, for example, when there
is at least one $g \in G^\textbf{k}$ that commutes with $a_0$ and
has a \emph{real} matrix representative.  With the new basis system:

\begin{equation}
\label{transformed_basis} \psi'=\psi+\phi,\hspace{6 pt}
\psi''=i(\psi-\phi)\end{equation}

the \textit{corep} decomposes into two identical \textit{coreps}:

\begin{eqnarray}
\label{a_type_transform}
g\psi'=\Delta(g)\psi', \hspace{6 pt}a_0g\psi'=\Delta(a_0ga_0^{-1})\psi'\\
g\psi''=\Delta(g)\psi'', \hspace{6
pt}a_0g\psi''=\Delta(a_0ga_0^{-1})\psi''\nonumber
\end{eqnarray}

The second \textit{corep} is often made antisymmetric with respect
to the antiunitary operators by omitting the imaginary unity in
the construction of $\psi''$. Clearly, the two resulting
\textit{coreps} $D_+$ and $D_-$ remain identical.  If $N$ is not
the identity, Eq. \ref{transformed_basis} must be generalized to:

\begin{equation}
\label{general_transformed_basis} \psi'=\psi+N^*\phi,\hspace{6 pt}
\psi''=i(\psi-N^*\phi)\end{equation}

All the examples in Section \ref{section: examples} belong to this "Case a".

\item[b \hspace{6 pt}] $\Delta(g)=N\overline{\Delta}(g)N^{-1}$ \emph{but}
$NN^*=-\Delta(a_0^2)$.  In this case, which is comparatively rare
for centrosymmetric groups, Eq. \ref{primitive_matrix} cannot be
reduced to a diagonal form.  Instead, an appropriate transformation
is applied to convert the matrix representatives of antiunitary
operators into a block-antisymmetric matrix.

\item[c \hspace{6 pt}] $\Delta(g)$ and $\overline{\Delta}(g)$ are \emph{not}
equivalent.  In this case, we retain the form of Eq.
\ref{primitive_matrix} with the same basis. Note that
$\overline{\Delta}(g)$ must necessarily be equivalent to one of the
other \textit{irreps} in the list, say $\overline{\Delta}(g)=
N\Delta'(g)N^{-1}$.
\end{description}

Complete tables of the \textit{coreps} for all crystallographic
space groups, as well as of the "auxiliary matrices" $N$ (therein
called $\beta$) are contained in Ref. \cite{Kovalev_1993}.  From
these tables, one can readily construct the \textit{corep} matrices
and the new basis vectors.

\end{document}